\documentclass[aip, amsmath,amssymb, reprint]{revtex4-1}
\usepackage{dcolumn}
\usepackage{color}
\usepackage{graphicx}
\usepackage{bm}
\usepackage{soul}
\usepackage[utf8]{inputenc}
\usepackage[T1]{fontenc}
\usepackage{mathptmx}
\usepackage{natbib}
\usepackage{upgreek}
\usepackage[normalem]{ulem}

\definecolor{OliveGreen}{rgb}{0,0.6,0}
\definecolor{NavyBlue}{rgb}{0.0, 0.07, 0.8}
\definecolor{ModeBeige}{rgb}{0.59, 0.44, 0.09}
\definecolor{MordantRed}{rgb}{1, 0.0, 0.0}
\definecolor{Orange}{rgb}{1.0, 0.5, 0.0}
\definecolor{SkyMagenta}{rgb}{0.81, 0.44, 0.69}
\definecolor{BlueGreen}{rgb}{0.0, 0.7, 0.7}
\definecolor{BlueViolet}{rgb}{0.54, 0.17, 0.89}

\usepackage{amsmath}
\usepackage{amsfonts}
\usepackage{amssymb}

\newcommand{\titre}{Thermoelectric and magneto-transport characteristics of interconnected networks of ferromagnetic nanowires and nanotubes}

\begin{document}

\preprint{AIP/123-QED}

\title{\titre}
\author{Tristan da C\^{a}mara Santa Clara Gomes}
\affiliation{Institute of Condensed Matter and Nanosciences, Universit\'{e} catholique de Louvain, Place Croix du Sud 1, 1348 Louvain-la-Neuve, Belgium}
\author{Nicolas Marchal}
\affiliation{Institute of Condensed Matter and Nanosciences, Universit\'{e} catholique de Louvain, Place Croix du Sud 1, 1348 Louvain-la-Neuve, Belgium}
\author{Joaqu\'{i}n de la Torre Medina}
\affiliation{Instituto de Investigaciones en Materiales/Unidad Morelia, Universidad Nacional Autónoma de México, Morelia, Mexico}
\author{Flavio Abreu Araujo}
\affiliation{Institute of Condensed Matter and Nanosciences, Universit\'{e} catholique de Louvain, Place Croix du Sud 1, 1348 Louvain-la-Neuve, Belgium}
\author{Luc Piraux}
\email{luc.piraux@uclouvain.be}
\affiliation{Institute of Condensed Matter and Nanosciences, Universit\'{e} catholique de Louvain, Place Croix du Sud 1, 1348 Louvain-la-Neuve, Belgium}


\begin{abstract}
Macroscopic-scale nanostructures, situated at the interface of nanostructures and bulk materials, hold significant promise in the realm of thermoelectric materials. Nanostructuring presents a compelling avenue for enhancing material thermoelectric performance as well as unlocking intriguing nanoscale phenomena, including spin-dependent thermoelectric effects. This is achieved while preserving high power output capabilities and ease of measurements related to the overall macroscopic dimensions. Within this framework, the recently developed three-dimensional interconnected nanowire and nanotube networks, integrated into a flexible polymer membrane, emerge as promising candidates for macroscopic nanostructures. The flexibility of these composites also paves the way for advances in the burgeoning field of flexible thermoelectrics. In this study, we demonstrate that the three-dimensional nanowire networks made of ferromagnetic metals maintain the intrinsic bulk thermoelectric power of their bulk constituent even for a diameter reduced to approximately 23 nm. Furthermore, we showcase the pioneering magneto-thermoelectric measurements of three-dimensional interconnected nickel nanotube networks. These macroscopic materials, comprising interconnected nanotubes, enable the development of large-area devices that exhibit efficient thermoelectric performance, while their nanoscale tubular structures provide distinctive magneto-transport properties. This research represents a significant step toward harnessing the potential of macroscopic nanostructured materials in the field of thermoelectrics.
\end{abstract}

\maketitle

\begin{figure*}[ht!]
\includegraphics[scale=0.8]{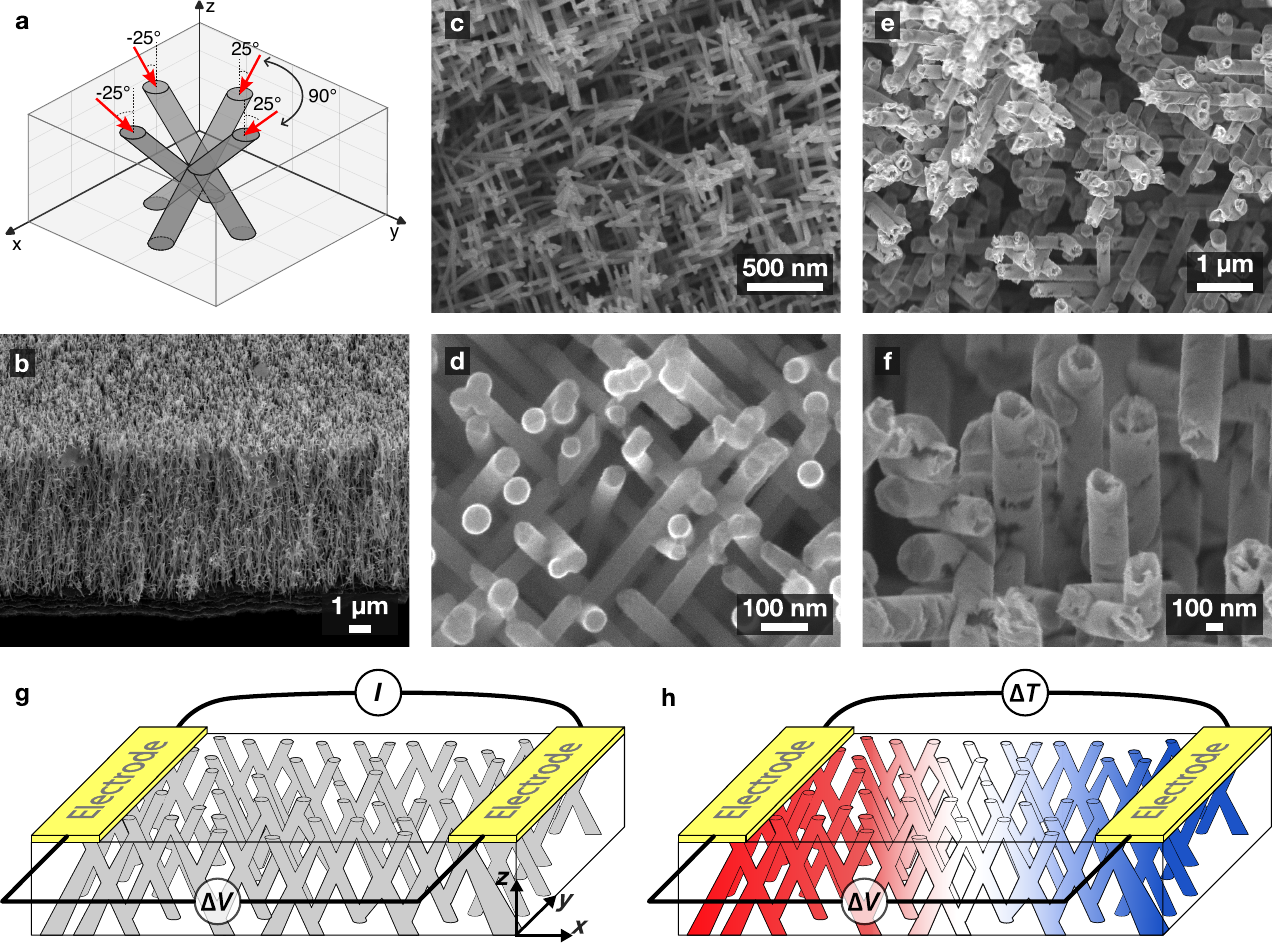}
\caption{\textbf{Three-dimensional interconnected networks and transport measurement configurations.} \textbf{a}, Schematic representation of the four irradiation directions for the track etching process, leading to 3-D networks of crossed nanopores. The 4 irradiation directions, successively offset by an angle of $90^\circ$ in the $x$-$y$ plane, form an angle of $25^\circ$ with the $z$-direction. \textbf{b-f}, Scanning electron microscope (SEM) images of self-supported and mechanically stable networks made of interconnected Ni nanowires and nanotubes. \textbf{b}, Titled view of the network film exhibiting a packing density of 20\%. \textbf{c-d}, SEM images of interconnected Ni nanowires with a diameter of 23 (\textbf{c}) and 45 nm (\textbf{d}).  \textbf{e-f}, SEM images of interconnected Ni nanotubes with a diameter of 230 nm and a tube thickness of approximately 45 nm. \textbf{g-h}, Illustrations of the device configurations employed for consecutive measurements of resistance and the Seebeck coefficient. In \textbf{(g)}, a voltage differential $\Delta V$ is induced by the injected current $I$ between the electrodes, while in \textbf{(h)}, it arises from a heat flow generated by a temperature difference $\Delta T$ between the two metallic electrodes, monitored by a thermocouple. Electrical and heat currents flow overall in the $x$-direction, while being locally restricted along the nanowire/nanotube axis, and an external magnetic field can be applied within any direction of the $y$-$z$ plane.}
\label{Fig1}
\end{figure*}

Macroscopic three-dimensional (3-D) nanostructures with complex architectures are attracting significant attention for their ability to exploit nanoscale phenomena within an efficient macroscopic configuration. This ability is particularly valuable in the field of thermoelectricity, where 3-D nanostructures serve as an ideal pathway to explore nanoscale phenomena while maintaining a substantial output power due to its macroscopic dimensions. Notably, nanostructuring has been proposed as a means to enhance the thermoelectric figure of merit ($ZT$), defined as $ZT = S^2  T  \sigma / \kappa$, where $S$ represents the Seebeck coefficient, $T$ is the temperature, $\sigma$ stands for electrical conductivity, and $\kappa$ is thermal conductivity. This enhancement can be achieved by increasing the thermopower due to quantum confinement and/or reduce lattice thermal conductivity by enhancing phonon scattering \cite{Lin2003, Abramson2004, Zhao2005, Dresselhaus2007, Bux2010, Fang2013, Hung2019, Sanad2020, Wu2022, yazawa2021thermoelectric}. Additionally, at the nanoscale, intriguing phenomena emerge, including spin-dependent thermoelectric effects. The emerging field of spin caloritronics harnesses these phenomena and is expected to play a pivotal role in the development of next-generation thermoelectric devices \cite{He2017, Uchida2010_Nature, Bauer2012, Boona2014, He2018, Walter2011}. In this context, the recent developments in the synthesis of 3-D networks made of interconnected magnetic nanowires (NWs) and nanotubes (NTs) offer perspectives for various nanodevices and nano-electronics \cite{Rauber2011, Hrkac2011, Parkin2008, Wang2012, Kwon2012, Araujo2015, Antohe2016, Piraux2016, Yershov2016,  Piraux2020, Torre-Medina2018, Camara-Santa-Clara-Gomes2019, Pacheco2017, Bhattacharya2022}. Their unique architecture offers excellent mechanical stability of the self-standing nanostructures, a high surface over volume ratio and lightness to the network films. The branching architecture provides excellent electrical and thermal connectivity, enabling reliable transport measurements within the plane of the network films \cite{Camara-Santa-Clara-Gomes2016_JAP, Torre-Medina2018, Camara-Santa-Clara-Gomes2019}. Additionally, the structural flexibility and mechanical robustness of 3-D interconnected network grown in flexible polymer template, together with the ability the integration of p-n junctions within a single flexible polymer membrane, are particularly valuable for the development of light-weight and flexible devices \cite{Camara-Santa-Clara-Gomes2019, Abreu-Araujo2019, Camara-Santa-Clara-Gomes2021_new, JPD2022, nano2023}. These prospects have recently attracted attention because they open avenues for applications in magneto-electronics \cite{Melzer2015, Makarov2016, Wang2016, Liu2013} and wearable thermoelectric devices \cite{Fan2021, Zhang2021, Bahk2015, Du2018, Lee2019, Masoumi2022, Petsagkourakis2018, Sun2022, Cao2023}. Recent studies have shown that ferromagnetic nanowire arrays exhibit very large thermoelectric power factors while both p-type and n-type thermoelectric materials can be obtained \cite{JPD2022, Camara-Santa-Clara-Gomes2019_2, Nico2023}. Additionally, fine-tuning of the thermopower can be achieved by controlling the alloying composition \cite{Camara-Santa-Clara-Gomes2019_2, Marchal2021}. As a result, these flexible 3-D structures hold great promise for active cooling \cite{nano2023}, magnetically activated thermoelectric switches \cite{Camara-Santa-Clara-Gomes2021_new}, thermal management and logic computing \cite{Camara-Santa-Clara-Gomes2021_new, nano2023}. Therefore, it is interesting to study the evolution of thermoelectric performance for very small diameter magnetic nanowires and to explore the evolution of thermopower in complex architectures made of interconnected Ni nanotubes that can be electrochemically fabricated in these nanoporous templates \cite{Torre-Medina2018, Piraux2020, Camara-Santa-Clara-Gomes2021}. For example, recent studies carried out on 3-D arrays of Bi NWs \cite{Wagner2021, Luc2023} showed a gradual decrease in thermopower because of an increasing contribution of surface charge carriers as the diameter decreases.

In this study, we explore the influence of dimensions and morphology on the thermoelectric and magneto-transport properties of interconnected networks consisting of ferromagnetic NW and NT networks. These 3-D nanostructures are fabricated through direct electrochemical deposition into track-etched polymeric membranes comprising a 3-D nanochannel network (see Supplementary Information Fig.~S1 for details). This synthesis method has proved to be a versatile and reliable method for fabricating large-area 3-D NW networks, offering control over dimensions, morphology, density of NWs or NTs, and material composition to create diverse complex 3-D networks of high aspect-ratio nanostructures \cite{Rauber2011, Araujo2015, Piraux2020, Torre-Medina2018, Camara-Santa-Clara-Gomes2021}. The 22 $\upmu$m thick track-etched polycarbonate (PC) template are obtained through four successive irradiation steps at angles of $\pm$25$^\circ$ ($\pm$5$^\circ$) with respect to the out-of-plane direction of the membrane along two perpendicular direction within the plane of the membrane, as illustrated in Fig.~\ref{Fig1}a.. Subsequent wet etching with NaOH was employed to define the cylindrical nanopores \cite{Ferain2003}, creating the intricate 3-D network of nanopores with precise control over the pore diameters $d$, ranging from 23 to 230 nm. Interconnected network of Ni \cite{Camara-Santa-Clara-Gomes2016_Nanoscale}, Co \cite{Camara-Santa-Clara-Gomes2016_JAP, Camara-Santa-Clara-Gomes2016_Nanoscale}, Fe \cite{Nico2023} and permalloy (Py) \cite{Marchal2020} NWs are grown within the crossed nanopore networks via electrodeposition from home-made electrolytes, using a metallic Cr/Cu bilayer on one side of the template as the cathode during the process. By adjusting the pH value of the Co-based electrolyte to either 2 or 5, we induce the formation of a dominant face-centered cubic (fcc) or hexagonal close-packed (hcp) crystalline structure, respectively \cite{Camara-Santa-Clara-Gomes2016_JAP, Camara-Santa-Clara-Gomes2016_Nanoscale}. Interconnected Ni NTs are obtained through the electrochemical dealloying method \cite{Torre-Medina2018, Camara-Santa-Clara-Gomes2021}. It involves two steps. First a Cu/Ni - core/shell structure is grown during the reduction step from a Cr/Au cathode. Secondly, the Cu core is removed during the oxidation step. By adjusting the reduction potential used, the NT thickness can be finely tuned, as demonstrated in ref. \cite{Torre-Medina2018}. In this work, interconnected NTs with wall thickness $t \approx$ 45 nm are considered. For all samples, the template have been partially filled with either NWs or NTs, typically between 10 to 15 $\upmu$m.

Figures~\ref{Fig1}b-f present scanning electron microscope (SEM) images of the NW and NT networks after the complete removal of the cathode and PC template. Notably, the NW and NT networks closely replicate the inverse geometry of the template, and they are found to be self-supported and mechanically stable. In Figure~\ref{Fig1}b, a tilted view of a NW network showcases the uniqueness of the nanostructure: it forms a single macroscopic film composed of physically interconnected nanofibers. These interconnections ensure excellent electrical and thermal connectivity within the network film. Consequently, by locally etching the metallic cathode used for electrodeposition to create a two-probe electrode configuration, as shown in Figs.~\ref{Fig1}g-h, it becomes possible to reliably measure the resistance and Seebeck coefficient along the macroscopic in-plane direction of the network films \cite{Camara-Santa-Clara-Gomes2016_JAP, Camara-Santa-Clara-Gomes2016_Nanoscale, Camara-Santa-Clara-Gomes2019, Abreu-Araujo2019}. The etched area of the samples measures approximately 2 mm in length and 2 to 5 mm in width. Resistance measurements are conducted by applying an electrical current between the two electrodes and recording the resulting voltage differential $\Delta V$ (see Fig.~\ref{Fig1}g). The electrical current predominantly flows along the $x$-direction within the plane of the network film, while being locally restricted along the NWs or NTs axes. The current amplitude was set between 10 $\upmu$A and 1 mA to ensure a drop voltage in the mV range and an input power of less than $\upmu$W. The Seebeck coefficient is determined by inducing a temperature difference $\Delta T$ between the two electrodes, which is achieved by connecting one electrode to a resistive heater while maintaining the other at the desired temperature, and recording the resulting voltage differential $\Delta V$ (see Fig.~\ref{Fig1}h). Typical temperature differences of 1 K are applied, which are recorded by a small-diameter type E differential thermocouple. Similarly, the thermal current flows primarily along the $x$-direction, with local restrictions along the NWs or NTs axes. To investigate the magnetic dependence of resistance and the Seebeck coefficient, we sweep an external magnetic field between $\pm$8.5 kOe in either in-plane (IP, $y$-direction) or out-of-plane (OOP, $z$-direction) configurations. The thermoelectric measurements are performed in the temperature range from 100 K to 300 K.

\begin{figure*}[ht]
\centering
\includegraphics[scale=0.8]{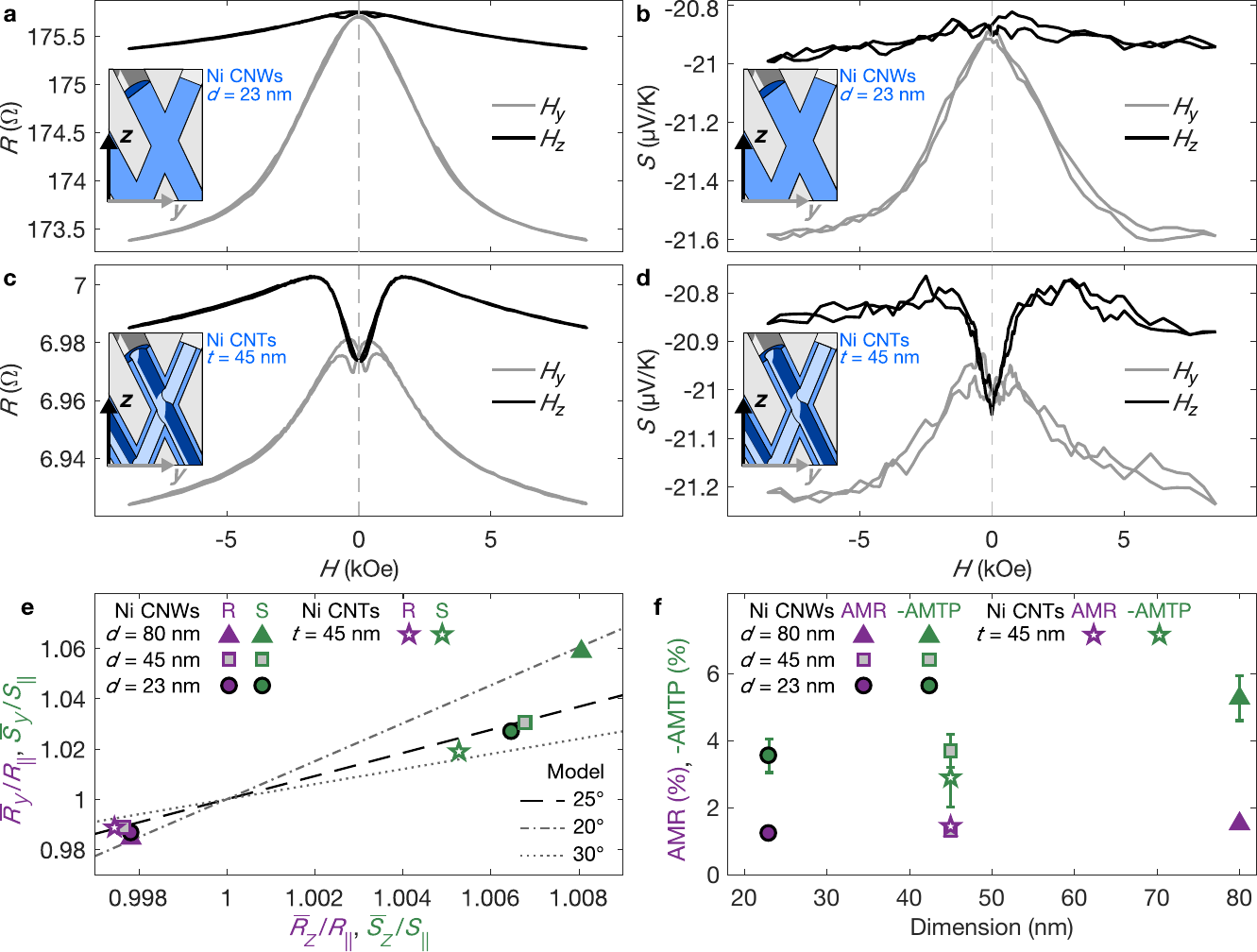}
\caption{\textbf{Magnetoresistance and magnetothermopower characteristics of crossed Ni nanowires and nanotubes.} \textbf{a-b}, Variation of the resistance ($R$, \textbf{a}) and Seebeck coefficient ($S$, \textbf{b}) measured on Ni crossed nanowire networks with a diameter of 23 nm, while applying a magnetic field in both the in-plane ($y$-direction) and out-of-plane ($z$-direction) orientations. \textbf{(c-d)} Corresponding magnetoresistance (\textbf{c}) and magneto-thermopower (\textbf{d}) characteristics obtained for Ni crossed nanotube networks with a diameter of 230 nm and a tube thickness of approximately 45 nm. \textbf{(e)} Relation between the resistance and the Seebeck coefficient data measured at saturation in both the $y$ and $z$ directions, normalized by the values in the parallel state given by Equation \ref{Eq1}. \textbf{(f)} Anisotropic magnetoresistance (AMR) and magneto-thermopower ($-$AMTP) as a function of their respective dimensions (nanowire diameter d and nanotube thickness), extracted using the analytical model. All the results shown in Figure 2 were obtained at room temperature.}
\label{Fig2}
\end{figure*}

Figure \ref{Fig2} shows the distinct magnetic behaviors that are observed for Ni NWs and NTs. In Figure \ref{Fig2}a, the magnetic dependence of resistance of a Ni NW network with a diameter $d$ of 23 nm is shown while applying the magnetic field in the $y$ and $z$ directions. The highest resistance state is reached at zero magnetic field, while the resistance decreases as the external magnetic field is increased, with a more pronounced reduction observed when the field is applied in the $y$ direction. This behavior can be attributed to the anisotropic magnetoresistance effect. Specifically, in the case of Ni NWs, the magnetostatic anisotropy dictates that magnetization aligns with the NW axes, parallel to the current paths, resulting in the highest resistance state \cite{Camara-Santa-Clara-Gomes2016_Nanoscale}. Under an applied magnetic field along the $y$ or $z$ directions, the magnetisation is rotated away from the NW axes, resulting in a decrease in resistance. Because the NW axes make an angle of 25$^\circ$ with the $z$ direction, the reduction in resistance is less significant when the magnetic field is applied along the $z$ direction compared to the $y$ direction. In Figure \ref{Fig2}b, we display the analogous variation of the Seebeck coefficient with the external magnetic field, which displays a similar shape. The measured Seebeck coefficient value of approximately $-21$ $\upmu$V/K is close to the reported bulk Ni value \cite{Rowe1995}, despite the 23 nm diameter of the NWs. Figures \ref{Fig2}c-d show the magnetic variation of the resistance and the Seebeck coefficient measured by sweeping an external magnetic field along the $y$ and $z$ directions of a Ni NT network with tube thickness $t$ of about 45 nm. We observe similar behaviors for large external magnetic fields induced by the saturation of magnetization along the $y$ and $z$ directions. However, contrasting behavior is observed close to zero magnetic field. As already observed in ref. \cite{Torre-Medina2018}, the NT configuration induce the formation of large number of vortex domain walls when the magnetic field is reduced to zero, leading to a large volume of the magnetisation that is not parallel to the NT axes and thus contributing to a smaller resistance state (see Fig \ref{Fig2}c and Fig. S2 in the Supplementary Information for the hysteresis loops). This effect also appears in the magnetic variation of the Seebeck coefficient (see Fig \ref{Fig2}d). These results clearly indicate the close relationship between the resistance and Seebeck coefficient in our Ni NW and NT networks.

Considering a given NW or NT segment with a specified orientation $\theta$ relative to the $z$-direction, its electrical resistivity depends on the relative orientation between the electrical current flow, restricted along the segment, and its magnetisation, that is $R_{\perp} + (R_{\parallel} - R_{\perp}) \cos^2{\theta}$, where $R_{\perp}$ and $R_{\parallel}$ represent the resistance values for configurations in which the current and magnetization are perpendicular and parallel to each other, respectively \cite{Camara-Santa-Clara-Gomes2016_JAP, Camara-Santa-Clara-Gomes2016_Nanoscale}. Because the NWs and NTs make an average angle of 25$^\circ$ with respect to the $z$ direction (with an estimated variation of $\pm$5$^\circ$ from the irradiation process of the 3-D porous polymer template), we can estimate the resistivity at saturation along the $z$ and $y$ directions as follows: $\bar{R}_{z} = \cos^2(25^\circ) R_{\parallel} + (1-\cos^2(25^\circ)) R_{\perp}$
and $\bar{R}_{y} = (1-\cos^2(25^\circ)) R_{\parallel} + \cos^2(25^\circ) R_{\perp}$. By taking the value $R_{\parallel}$ as the maximum resistance state reached during the magneto-resistance measurements, and the values of $\bar{R}_{z}$ and $\bar{R}_{y}$ as the resistance values measured at $H =$ 8.5 kOe along the $y$ and $z$ directions,  we can make a simple analytical estimation of the unknown value of $R_{\perp}$. The AMR ratio is then given by AMR $= \frac{R_{\parallel} - R_{\perp}}{1/3 R_{\parallel} + 2/3 R_{\perp}}$. The same approach is applied to extract the anisotropic magneto-thermopower ratio AMTP $= \frac{S_{\parallel} - S_{\perp}}{1/3 S_{\parallel} + 2/3 S_{\perp}}$, where $S_{\parallel}$ is estimated as the maximal value of the Seebeck coefficient, while $S_{\perp}$ is estimated from the measured value of the Seebeck coefficient at $H =$ 8.5 kOe in the $y$ and $z$ directions $\bar{S}_{z}$ and $\bar{S}_{y}$. It should be noted that our model assumes that the highest resistance state corresponds to the maximum of the $R(H)$ and $S(H)$ curves and that saturation of the magnetization is reached at $\pm8.5$ kOe. As the measurements show that the saturated state is not completely reached at the highest measured field, and as a slight deviation between current and magnetization can be expected in the crossing zones of the nanocylinders, the values of AMR and AMTP reported in Fig. \ref{Fig2}f are slightly underestimated and constitute lower limits for these ratios. This analytical method also allows to extract the expected relation between $\bar{X}_{y}$ and $\bar{X}_{z}$ as
\begin{equation}
\bar{X}_{y} = \frac{\cos^2(25^\circ) \bar{X}_{z} + (1-2\cos^2(25^\circ))}{1-\cos^2(25^\circ)}\text{.}
\label{Eq1}
\end{equation}
using either $\bar{X}_{y} = \bar{R}_{y}/R_{\parallel}$ and $\bar{X}_{z} = \bar{R}_{z}/R_{\parallel}$ or $\bar{X}_{y} = \bar{S}_{y}/S_{\parallel}$ and $\bar{X}_{z} = \bar{S}_{z}/S_{\parallel}$. Figure \ref{Fig2}e shows the comparison between Equation \ref{Eq1} and the experimental dependence of $\bar{R}_{y}/R_{\parallel}$ vs. $\bar{R}_{z}/R_{\parallel}$ and  $\bar{S}_{y}/S_{\parallel}$ vs. $\bar{S}_{z}/S_{\parallel}$ for Ni NW and NT networks. The experimental data lie within the dashed and dotted grey lines that account for the irradiation angle variation of $\pm$ 5$^{\circ}$, which validates the simple analytical model for both nanostructures. The estimated values of the AMR and AMTP ratios are presented in Fig. \ref{Fig2}f. We observe an AMR ratio just under  2\%, which is close to the values for Ni nanowire networks \cite{Camara-Santa-Clara-Gomes2016_Nanoscale, Torre-Medina2018, Camara-Santa-Clara-Gomes2019_2, Camara-Santa-Clara-Gomes2021} and individual Ni nanowires \cite{Bohnert2013, Niemann2016} studied previously. This value is slightly lower than the AMR ratio reported for bulk Ni ($\approx2$\%) \cite{McGuire1975} due to additional diffusion mechanisms related to structural defects and reduced lateral dimensions of the electrodeposited NWs. The magnitude of the AMTP ratio is 2 to 3 times higher than the corresponding AMR effect, which is in agreement with previous results obtained on bulk Ni \cite{Uchida2018} and single Ni nanowires of diameter $>$ 100 nm \cite{Bohnert2013, Niemann2016},  which report an MTEP ratio around 3\% to 6\%. It is worth noting that the AMTP ratios are negative, a consequence of the negative sign of the Seebeck coefficient in Ni.

\begin{figure*}[ht]
\centering
\includegraphics[scale=0.8]{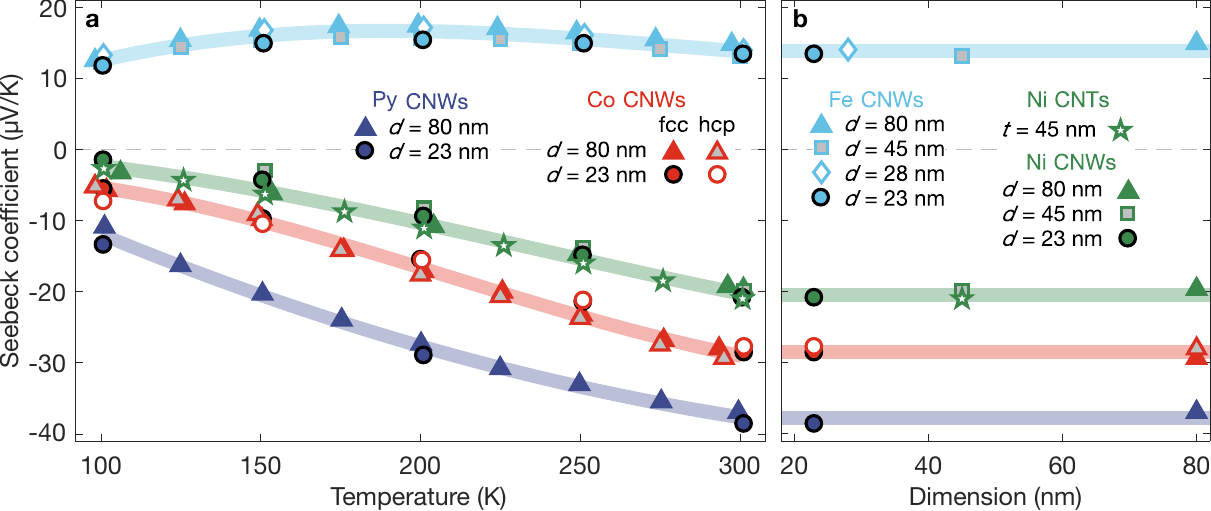}
\caption{\textbf{Evolution of the Seebeck coefficient of ferromagnetic interconnected NW and NT networks as a function of temperature for different NW diameters and specific NT wall width.} \textbf{a}, Seebeck coefficient of ferromagnetic crossed nanowires (NWs) with varying diameters ($d$) and Ni crossed nanotubes (NTs) with wall thickness ($t$) of 45 nm. The measurements are reported for Co nanowires exhibiting both dominant face-centered cubic (fcc) and hexagonal close-packed (hcp) crystalline structures. \textbf{b}, Room temperature thermopower of the samples in \textbf{a} as a function of their respective dimensions (nanowire diameter $d$ and nanotube thickness $t$).}
\label{Fig3}
\end{figure*}

We now shift our discussion to the influence of dimensions on the thermopower of interconnected NW and NT networks comprising various ferromagnetic metals and alloys. In Figure \ref{Fig3}a, the thermopower data obtained for Fe, Co, Ni, and Py NWs with diameters $d$ ranging from 23 nm to 80 nm, as well as Ni NT with tube thickness $t$ of $\sim$45 nm, is presented for temperatures spanning from 100 K to 300 K. Very slight variation in thermopower is observed for all materials within the considered range of dimensions, which can be attributed to the experimental measurement uncertainties. Regarding the Co NWs, the results in Fig.\ref{Fig3}a indicate that the thermopower is not modified following the change of crystalline structure due to electrochemical deposition at pH 2 and 5. Indeed, as shown previously \cite{Camara-Santa-Clara-Gomes2016_JAP, Camara-Santa-Clara-Gomes2016_Nanoscale, Camara-Santa-Clara-Gomes2019_2, Camara-Santa-Clara-Gomes2020, Camara-Santa-Clara-Gomes2021}, an fcc-type structure characterizes Co NWs made at lower pH while Co NWs synthesized at pH 5 present an hcp structure with the c-axis preferentially oriented perpendicular to the nanowire axis.  For the latter, this results in a modification of the magnetic anisotropy due to a strong transverse component in the magneto-crystalline anisotropy linked to the hcp structure. As previously shown in refs \cite{Camara-Santa-Clara-Gomes2016_JAP, Camara-Santa-Clara-Gomes2016_Nanoscale, Camara-Santa-Clara-Gomes2019_2, Camara-Santa-Clara-Gomes2020, Camara-Santa-Clara-Gomes2021}, this results in a slight decrease of the electrical resistance near zero magnetic field, which is reflected in the field evolution of the Seebeck coefficient. Interestingly, we observe a similar temperature variation of the Seebeck coefficient in the Ni NT network when compared to the Ni NW networks, which suggests that the complex nanostructure with high surface-to-volume ratio has no significant impact on the relatively high Seebeck coefficient of Ni. Unlike the other materials, Fe NWs exhibit a significantly large and positive Seebeck coefficient, which has been recently reported in 3-D NW networks and attributed to a pronounced magnon drag effect \cite{Nico2023}. Figure \ref{Fig3}b shows the Seebeck coefficient measured at room temperature on the various NWs and NTs as a function of the dimension ($d$ or $t$). This highlights the absence of dimension effect on the thermoelectric properties of ferromagnetic nanofiber networks. It should be noted that the Seebeck coefficient values measured were found to be very close to that of the bulk constituent. This indicates that macroscopic interconnected networks of nanofibers keep the thermoelectric properties of the bulk constituent even for nanofiber dimensions down to 23 nm.

In conclusion, we have reliably measured the thermoelectric and magneto-transport properties along the in-plane direction of macroscopic interconnected network films made of ferromagnetic nanowires (with diameter ranging from 80 to 23 nm) and nanotubes (with tube thickness of approximately 45 nm). Our observations reveal distinct magneto-transport properties in Ni nanotubes compared to Ni nanowires, which can be attributed to the nucleation of a large number of vortex-like domain walls when the magnetic field is reduced to zero. Furthermore, our results indicate that the thermopower of the nanowire network remains unchanged from the bulk value over a temperature range from 100 K to 300 K, and this correspondence holds for dimensions down to 23 nm. The same applies to the thermopower of nanotube structures with high surface-to-volume ratio. These structures have great potential for the development of thermoelectric devices, as they combine high flexibility and good shapeability, macroscopic dimensions and excellent interconnectivity of the ferromagnetic nanocylinder network.\\

\noindent
\textbf{Supplementary Material}\\
The Supplementary Material provides additional details regarding the fabrication process of 3-D interconnected nanowire and nanotube networks, as well as a comparison between the magnetization and magnetoresistance curves of 3-D interconnected Ni nanowire and nanotube networks.\\

\noindent
\textbf{Acknowledgments}\\
Financial support was provided by the Belgian Fund for Scientific Research (F.R.S.-FNRS) and F.A.A. is a Research Associate of the F.R.S.-FNRS. JdlTM thanks CONAHCYT for finantial support through project A1-S-9588. The authors would like to thank Dr. E. Ferain and the it4ip Company for supplying polycarbonate membranes.\\

\noindent
\textbf{REFERENCES}
\bibliographystyle{naturemag}


\end{document}